\documentstyle[12pt,epsf]{article}

\catcode`@=12
\newcommand{\mrm}[1]{\;\mbox{\rm #1}}
\newcommand{\beq}{\begin{equation}}
\newcommand{\eeq}{\end{equation}}
\newcommand{\nn}{\nonumber}
\newcommand{\bea}{\begin{eqnarray}}
\newcommand{\eea}{\end{eqnarray}}

\newcommand{\rfn}[1]{(\ref{#1})}
\newcommand{\Eq}[1]{Eq.~(\ref{#1})}
 
\newcommand{\ea}{{\it et al.}}

\newcommand{\ie}{{\it i.e.}}
\newcommand{\eg}{{\it e.g.}}
 
\newcommand{\np}[1]{{ Nucl. Phys. }{\bf #1}}
\newcommand{\pl}[1]{{ Phys. Lett. }{\bf #1}}
\newcommand{\pr}[1]{{ Phys. Rev. }{\bf #1}}
\newcommand{\prl}[1]{{ Phys. Rev. Lett. }{\bf #1}}

\newcommand{\epj}[1]{{ Eur. Phys. J. }{\bf #1}}

\def\lsim{\mathrel{\vcenter{\hbox{$<$}\nointerlineskip\hbox{$\sim$}}}}
\def\gsim{\mathrel{\vcenter{\hbox{$>$}\nointerlineskip\hbox{$\sim$}}}}         

\begin{document}
\thispagestyle{empty}
\begin{flushright} UCRHEP-T295\\ December 2000\
\end{flushright}
\vspace{0.3in}
\begin{center}
{\Large \bf  Phenomenology of the Neutrino-Mass-Giving \\ Higgs
Triplet and the Low-Energy Seesaw \\ Violation of Lepton Number\\}
\vspace{0.7in}
{\bf Ernest Ma$^1$, Martti Raidal$^{1,2}$, and Utpal Sarkar$^3$\\}
\vspace{0.1in}
{\sl $^1$ Physics Department, University of California, Riverside, 
California 92521, USA\\}
{\sl $^2$ National Institute of Chemical Physics and Biophysics, 
Tallinn 10143, Estonia\\}
{\sl $^3$ Physical Research Laboratory, Ahmedabad 380 009, India\\}
\vspace{0.8in}
\end{center}
\begin{abstract}
Small realistic Majorana neutrino masses can be generated via a Higgs 
triplet $(\xi^{++}, \xi^+, \xi^0)$ without having energy scales larger 
than $M_*={\cal O}(1)$ TeV in the theory.  The large effective mass scale 
$\Lambda$ in the well-known seesaw neutrino-mass operator $\Lambda^{-1} 
(LL\Phi\Phi)$ is naturally 
obtained with $\Lambda\sim M_*^2/\mu,$ where $\mu$ is a {\it small} scale of 
lepton-number violation.  In theories with large extra dimensions, the 
smallness of $\mu$ is naturally obtained by the mechanism of ``shining'' 
if the number of extra dimensions $n\ge 3.$  We study here the Higgs 
phenomenology of this model, where the spontaneous violation of lepton 
number is treated as an external source from extra dimensions.  The 
observable decays $\xi^{++} \to l_i^+l_j^+$ will determine directly the 
magnitudes of the $\{ij\}$ elements of the neutrino mass matrix. 
The decays $\xi^+ \to W^+ J^0$ and $\xi^0 \to Z J^0$, where $J^0$ is the 
massless Goldstone boson (Majoron), are also possible, but of special 
importance is the decay $\xi^0 \to J^0 J^0$ which provides stringent 
constraints on the allowed parameter space of this model.  Based on the 
current neutrino data, we also predict observable rates of $\mu-e$ 
conversion in nuclei.
\end{abstract}

\newpage
\baselineskip 24pt

\section{Introduction}

The idea that standard-model (SM) fields can be localized on a brane
has greatly changed our approaches to quantum gravity and extra 
dimensions.  It allows the fundamental scale of quantum gravity
to be as low as ${\cal O}(1)$ TeV \cite{ex}, thus  providing distinctive 
experimental signatures of extra dimensions at future colliders \cite{excoll}. 
These include the radiation of gravitons into extra dimensions and the 
exchange of graviton Kaluza-Klein towers which modify SM neutral-current 
processes.

An important development in these TeV-scale extra-dimensional theories is 
their implication on new approaches to low-energy model building.  They 
allow us to consider new possibilities in addressing current topics of 
interest in particle physics.  Perhaps the most important of them is the 
question of nonzero neutrino masses, as indicated by the atmospheric and solar 
neutrino anomalies \cite{neutrino}.  In extra-dimensional theories, 
there is a possibility that neutral gauge singlet particles
are not confined to our 3-brane and  propagate in the full volume of the 
theory,  similarly to the gravitons.  The first idea of explaining the 
smallness of neutrino masses in extra-dimensional theories was to 
introduce singlet neutrinos in the bulk \cite{exnu,exnu2}. These become the 
right-handed partners of the observed left-handed neutrinos, but with small 
Dirac masses, being suppressed by the large volume of the extra dimensions.

More recently, we have proposed an alternative scenario \cite{marasa,ml}
which obtains small {\it Majorana} neutrino masses instead.  Given the SM 
particle content, such neutrino masses are induced by a single effective 
operator \cite{wein}
\bea
\frac{1}{\Lambda}LL\Phi\Phi\,,
\label{one}
\eea
where $L$ and $\Phi$ are the SM lepton and Higgs-boson doublets 
respectively, and $\Lambda$ is the {\it effective} mass scale of 
the new physics; 
any extension of the SM to generate small Majorana neutrino masses is merely 
a particular realization of this effective operator 
provided the particle content of the model below the
electroweak scale remains the SM one \cite{ma}. 
The scale $\Lambda$ must be very large, ${\cal O}(10^{12}-10^{15})$
GeV, to account for realistically small neutrino masses.  It is usually 
associated with the scale of lepton-number violation, as in all models with 
the canonical seesaw mechanism \cite{seesaw} in which the small neutrino 
masses are inversely proportional to the very large right-handed Majorana 
neutrino masses, $\Lambda\sim M_N.$  However, this is just one specific 
solution which cannot be preferred over any other possible realization 
of Eq.~(1) from the viewpoint of low-energy  phenomenology . For example, if
\bea
\frac{1}{\Lambda}\sim\frac{\mu}{M^2}\,,
\label{newla}
\eea
where $M$ is the fundamental scale of new physics ($M\sim{\cal O}(1)$ TeV 
in our case) and $\mu$ is the {\it small} scale associated with the 
breaking of lepton number, we still achieve the same very large effective 
scale $\Lambda$ in \Eq{one} {\it without actually having the very large scale 
in the theory.}  Similar possibilities of generating small neutrino masses 
with a low-scale seesaw mechanism have also been considered recently in 
Ref.~\cite{v32}.  The crucial advantage of the low-scale seesaw mechanism 
\rfn{newla} over the canonical scenario is that the former allows
unambiguous tests of the neutrino-mass-giving mechanism at future colliders 
while the latter cannot be directly probed at terrestrial experiments.

In our recently proposed scenario \cite{marasa} 
(for first similar proposals see \cite{exnu}), we use the distant breaking 
\cite{distant} of lepton number to generate small Majorana neutrino masses 
through a scalar Higgs triplet \cite{masa} localized in our 3-brane.
The trilinear interaction of the triplet with the SM Higgs doublets
required for this mechanism is induced by the ``shining'' of a scalar
singlet which lives in the bulk and communicates the breaking of 
lepton number from another brane to our world.  The smallness of neutrino 
masses, or equivalently the smallness of the lepton-number breaking scale 
$\mu$ in \Eq{newla}, comes from the suppression of the Yukawa potential of 
the bulk singlet by the large separation of the two branes in the extra 
dimensions.  To achieve the desired suppression, the number of extra 
dimensions $n$ should be at least 3.  Notice that the structure of TeV-scale 
quantum gravity itself suggests naturally such a mechanism 
of breaking global quantum numbers because it contains ``our world''
on a 3-brane, the ``hidden sector'' on another brane not 
necessarily identical to our brane, and the ``messenger sector''
with particles living in the bulk.  The analogous mechanism for 
breaking the Peccei-Quinn symmetry has been considered in Ref. \cite{pq}.

The aim of the present paper is to work out details of our model proposed 
in Ref.~\cite{marasa}, concentrating on the collider and low-energy 
phenomenology predicted by this model.  In doing that, we first have to 
address the question of how to treat the field theory in our 3-brane 
consistently if $n$ extra dimensions are present at scales ${\cal O}(1)$ TeV. 
We show that this can be achieved by considering the 4-dimensional field 
theory in the presence of an external source. In this case the singlet bulk 
field decouples from the fields in our world, except for the massless Majoron 
which propagates also in the bulk and provides the only connection between 
these two sectors.

We first work out the Higgs phenomenology of the model and show that the 
decay of the neutral component of the Higgs triplet into two Majorons, i.e. 
$\xi^0 \to J^0 J^0,$  stringently constrains the pattern of its vacuum 
expectation values (VEVs).  Since the neutrino mass matrix in this model 
is uniquely fixed by the Yukawa couplings of the Higgs triplet to the 
leptons, the existing neutrino data may be used to determine the entries 
of this Yukawa matrix up to a normalization scale.  We can thus predict 
the rates of rare unobserved lepton-flavour violating processes both at 
low-energy experiments as well as at colliders.  Conversely, if the Higgs 
triplet is kinematically accessible at future colliders, the decay branching 
fractions of its doubly charged component into charged leptons, i.e. 
$\xi^{++}\to l^+_il^+_j$, will determine uniquely the relative magnitude of 
each element of the neutrino mass matrix.  Our neutrino-mass scenario is 
thus directly and unambiguously 
testable at colliders and can foretell the results of future
neutrino factories and long-baseline neutrino-oscillation experiments.

This paper is organized as follows. In Section 2 we introduce the
model. In Section 3 we study the Higgs potential of our model.  In Section 4 
we work out the details of the Higgs-boson phenomenology at colliders. In 
Section 5 we study the connection between neutrino-oscillation data and 
lepton-flavour violating processes.  Conclusions are in Section 6.

\section{The Triplet Model in the 
Presence of Extra Dimensions}

The fermion sector of our model is identical to that of the SM, containing 
the left-handed lepton doublets and right-handed charged-lepton singlets 
with the following $SU(2)_L\times U(1)_Y$ quantum numbers:
\bea
L_i=\left(\begin{array}{c} \nu_i \\ l_i\end{array}\right)_L\sim (2,-1/2)\,,
~~~~~~~~  l_{iR}\sim (1,-1)\,,
\eea
where $i=e,\,\mu,\,\tau.$
There are no right-handed neutrinos in the model.

The Higgs sector consists of the usual SM doublet 
\bea
\Phi =\left(\begin{array}{c} \phi^+ \\ \phi^0 \end{array}\right)\sim (2,1/2)\,,
\eea
and two additional scalar fields, a triplet $\xi$ and a singlet $\chi$:
\bea
\xi =\left(\begin{array}{cc} \xi^+/\sqrt{2} & \xi^{++}\\ 
 -\xi^0 & -\xi^+/\sqrt{2} \end{array}\right)\sim (3,1)\,,
~~~~~~~~~\chi=\chi^0\sim (1,0)\,.
\eea
The latter two fields carry lepton number $L=-2.$ The Higgs triplet $\xi$
is presented in the form of a $2\times 2$ matrix transforming under $SU(2)$
as $\xi \to U\xi U^\dagger.$ The triplet couples to leptons 
via the Yukawa interaction
\bea
{\cal L}_Y= f_{ij} L_i^T C^{-1}\, i\tau_2\, \xi\, L_j + h.c.\,.
\label{yuk}
\eea
If neutrinos obtain Majorana masses via \Eq{yuk}, lepton number is broken 
by two units. It is important to notice that the neutrino mass matrix in 
this case is proportional to a single Yukawa matrix $f_{ij}.$  Thus any 
inference from neutrino-oscillation data regarding the actual form of the 
neutrino mass matrix may now be tested in low-energy lepton-flavour 
violating processes as well as in collider experiments where $\xi$ may be 
produced and its decays observed.  The results of future neutrino factories 
and long-baseline neutrino-oscillation experiments may also be predicted.

If the lepton-number violation occurs spontaneously \cite{gero} via the VEV 
of the triplet in \Eq{yuk}, a massless Majoron will appear.  Based on 
searches for it at LEP via the invisible width of the $Z$, this model is 
ruled out.  However, if lepton number is violated explicitly \cite{scva} 
by the trilinear coupling $\mu \Phi^\dagger \xi \tilde \Phi$ as in 
Ref.~\cite{masa}, there is no (triplet) Majoron and no contradiction with 
present experimental data (for allowed Majoron models see, e.g., Ref. 
\cite{vallem}).


In our model as proposed in Ref.~\cite{marasa}, the SM fields together with 
$\xi$ are localized in our world (a 3-brane ${\cal P}$ at $y = 0$) and are 
blind to the extra space dimensions.  Lepton number is assumed to be 
conserved as far as these fields are concerned.  The Higgs singlet $\chi$ 
which carries lepton number is special because
\begin{itemize}
\item it propagates also in the bulk; 
\item it serves as a ``messenger'' which communicates the violation of
lepton number from another brane to our world through the large extra
dimensions. 
\end{itemize}
We assume the existence of a field $\eta$ which is localized in a distant 
3-brane (${\cal P}'$) situated at a point $y = y_*$ in the extra dimensions. 
It is a singlet under the standard model but has $L = 2$ and couples to 
$\chi$ (with $L=-2$).  When the field $\eta$ acquires a VEV, lepton number is 
broken maximally in the other brane.  It will then act as a point source for 
$L$ violation, and the field $\chi$ is the messenger which carries it to our 
wall (the interface between our brane and the bulk).  The ``shining'' of 
$\chi$ at all points in our world is the mechanism \cite{distant} which 
breaks lepton number and gives mass to the neutrinos.

At energies much below the fundamental scale $M_*$, the lepton-number 
violating effect will be suppressed by the distance between the source brane 
at ${\cal P}'$ and our brane at ${\cal P}$.  We assume that the source brane 
is situated at the farthest point in the extra dimensions so that $|y_*| = r$ 
is the radius of compactification and it is related to the fundamental scale 
$M_*$ and the reduced Planck scale ($M_P=2.4 \times 10^{18}$ GeV) by the 
relation
\begin{equation}
r^n M_*^{n + 2} \sim M_P^2  .
\label{rm}
\end{equation}
This explains why lepton number is only violated weakly in our world.

We assume here that the source brane has the same dimensional structure as 
our world and there are $n$ extra dimensions.  In our world (${\cal P}$) the 
field $\chi$ has only the lepton-number conserving interaction of the
form $\Phi(x)^\dagger \xi(x) \tilde \Phi(x) \chi^\dagger(x, y=0).$  In 
the other brane (${\cal P}'$) the field $\chi$ couples to the field 
$\eta$ through the interaction
\begin{equation}
{\cal S}_{other} = \int_{{\cal P}'} d^4 x' ~\mu^2~ \eta (x') \chi 
(x', y = y_*), 
\end{equation}
where $\mu$ is a mass parameter.  Lepton-number violation from $\langle \eta 
\rangle$ is carried by $\chi$ to our world through its ``shined'' value 
$\langle \chi \rangle$: 
\begin{equation}
\langle \chi (x, y = 0) \rangle = \Delta_n (r) \langle \eta (x, y = y_*) 
\rangle,
\end{equation}
where $\langle \eta \rangle$ acts as a point source, and $\Delta_n(r)$ is 
the Yukawa potential in $n$ transverse dimensions, i.e. \cite{distant}
\begin{equation}
\Delta_n (r)= {1 \over (2 \pi )^{n \over 2}
M_*^{n- 2}} ~\left( {m_\chi \over r} \right)^{n-2 \over 2}
~K_{n - 2 \over 2} \left( m_\chi r \right),
\end{equation}
$K$ being the modified Bessel function.  If the mass of the carrier 
field $\chi$ is large ($m_\chi r \gg 1$), it has the profile
\bea
\langle \chi \rangle \approx \displaystyle{ 
m_\chi^{n - 3 \over 2} \over 2 (2 \pi)^{n - 1 
\over 2} M_*^{ n - 2} } \displaystyle{e^{- 
m_\chi r} \over r^{n-1 \over 2} }.
\eea
The suppression here is exponential, hence the amount of lepton-number 
violation in our world is very small, but its precise value depends 
sensitively on $m_\chi$.  An interesting alternative is to have a light 
carrier field with a mass less than $1/r$.  However, it should be larger 
than about (0.1 mm)$^{-1}$, to be consistent with the present experimental 
data on gravitational interactions.

If $m_\chi r \ll 1$, $\Delta_n(r)$ is logarithmic for $n=2$ and $\langle 
\chi \rangle$ is not suppressed.  For $n>2$, the asymptotic form of the 
profile of $\chi$ is
\begin{equation}
\langle \chi \rangle \approx { \Gamma ( {n -2 \over 2} ) \over
4 \pi^{n \over 2} }{M_* \over (M_* r)^{n-2} },
\label{chivev}
\end{equation}
which is suitably small for large $r$. Because of the interaction 
$\Phi^\dagger \xi \tilde \Phi \chi^\dagger$ in our brane, the triplet 
$\xi$ will acquire a small effective VEV via tadpole diagrams as $\Phi$ 
and $\chi$ acquire VEVs. Thus small Majorana neutrino masses are generated via 
\Eq{yuk}.  Because the breakdown of lepton number in the distant
brane occurs spontaneously, there exists a (singlet) Majoron
which propagates also in the bulk and 
plays a very special role in our model.


To realize the above-described idea of neutrino masses (or any other
in which some particles propagate both on the 3-brane and also in the bulk), 
one has to answer a nontrivial question: i.e. {\it how to describe the 
four-dimensional field theory in our brane consistently if the large extra 
dimensions are present?} Clearly, in the full $4+n$-dimensional theory, the 
precise answer should be derived from a consistent theory of quantum gravity 
which is, however, not yet available.  On the other hand, the SM particles 
are confined on a 3-brane, and we do know how to treat them in the context 
of a field theory in four dimensions.  To take into account possible new 
physics effects from extra dimensions, we propose in this work three simple 
{\it ans\"{a}tze.}  We assume that:
\begin{itemize}
\item[(i)] independently of the origin of lepton-number violation, it is
communicated to our world solely via the small nonzero VEV of the 
field  $\chi;$ details of the physics in extra dimensions giving rise to it 
are irrelevant for our phenomenological approach; 
\item[(ii)] the value of the VEV of the singlet $\chi$ does not depend 
on the parameters in our 3-brane \ie, on other parameters of the model;
\item[(iii)] the Majoron is the massless field (in analogy to the graviton) 
which may cross from the bulk to the brane. 
\end{itemize}
It follows from these assumptions that the model dynamics in our world 
should be determined using the four-dimensional field theory in the presence 
of {\it an external source} (provided in our model by $\chi$ via the 
``shining'' mechanism.)  As we show in the next section, the way $\chi$ 
is being treated in our model is very different from what it would be if it 
were an ordinary singlet confined to our brane.

\section{Consistent Treatment of the Higgs Potential}

For our purpose it is convenient to express the bulk field $\chi$ as
\bea
\chi = {1 \over \sqrt 2} (\rho + z) e^{i\varphi}\,,
\eea
where $z/\sqrt 2 \equiv \langle\chi\rangle$ denotes the VEV of the field 
in our brane.  According to our assumptions (i) and (ii) in Section 2, the VEV 
$z$ should be regarded as a boundary condition, which is not
altered by  any other parameter of the model; all such effects
are already included in $z.$ In theories of extra dimensions, 
$z$ is induced by the ``shining'' mechanism and its numerical
value is small, $z<<<M_Z.$ The lepton-number transformations of $\rho$ 
and $\varphi$ under $U(1)_L$ are given by 
\bea
\rho\to \rho\,,~~~~ \varphi\to\varphi - 2x\,,
\eea
while the $U(1)_L$ transformations for neutrinos and the Higgs triplet read 
as usual:
\bea
\nu\to e^{ix} \nu\,, ~~~~ \xi\to e^{-2ix} \xi\,.
\eea
The self-interaction terms for the bulk scalar $\rho$ can now be 
expressed as
\bea
V(\chi)=\lambda_\chi\, z^2\rho^2 + 
\lambda_\chi\, z\,\rho^3 + \frac{1}{4}\lambda_\chi\,\rho^4\,
\eea
and the lepton-number conserving Higgs potential of the other fields as 
\begin{eqnarray}
V &=& m_0^2 \,\Phi^\dagger \Phi + m_\xi^2\, \mrm{Tr}[\xi^\dagger \xi] + 
{1 \over 2} \lambda_1 (\Phi^\dagger \Phi)^2 + 
{1 \over 2} \lambda_2 \mrm{Tr}[\xi^\dagger \xi]^2 + 
\lambda_3 (\Phi^\dagger \Phi)\mrm{Tr}[\xi^\dagger \xi] \nonumber \\ 
&& + \lambda_4 \mrm{Tr}[\xi^\dagger \xi^\dagger]\mrm{Tr}[\xi \xi]
+ \lambda_5 \Phi^\dagger \xi^\dagger \xi \Phi +
\left( {\lambda_0 z e^{-i\varphi} \over \sqrt 2} 
\Phi^\dagger \xi \tilde \Phi + h.c. \right) \,,
\label{V}
\end{eqnarray}
where $m_0^2 < 0$, but $m_\xi^2 > 0$.  Notice the presence of the VEV $z$ 
in the last term of \Eq{V}, which gives rise to the desired trilinear 
coupling of the Higgs doublets to the triplet.

In a similar fashion, let us express
\begin{equation}
\phi^0 = {1 \over \sqrt 2} (H + v) e^{i\theta}, ~~~~ 
\xi^0 = {1 \over \sqrt 2} (\zeta + u) e^{i\eta},
\end{equation}
where $v/\sqrt 2$ and $u/\sqrt 2$ are the vacuum expectation values of 
$\phi^0$ and $\xi^0$ respectively.  This way of writing allows a simple 
and consistent treatment of the massless Goldstone modes of the model.
Consider now only the neutral scalar fields $H$, $\zeta$, and the 
correctly normalized fields $v \theta$, $u \eta$, and $z \varphi$, then
\begin{eqnarray}
V_0 &=& {1 \over 2} m_0^2 (H+v)^2 + {1 \over 2} m_\xi^2 (\zeta + u)^2 + 
{1 \over 8} \lambda_1 (H+v)^4 + {1 \over 8} \lambda_2 (\zeta + u)^4 + 
{1 \over 4} \lambda_3 (H+v)^2 (\zeta + u)^2 \nonumber \\ &-& {\lambda_0 
z \over 2} (H+v)^2 (\zeta + u) \left[ 1 - {1 \over 2} (\varphi - \eta + 2 
\theta)^2 + ... \right].
\end{eqnarray}
The minimum of $V_0$ is determined by the first-derivative conditions
\begin{eqnarray}
m_0^2 + {1 \over 2} \lambda_1 v^2 + {1 \over 2} \lambda_3 u^2 - \lambda_0 z u 
&=& 0, \nn\\ 
u \left(m_\xi^2 + {1 \over 2} \lambda_2 u^2 + {1 \over 2} \lambda_3 
v^2 \right) - {1 \over 2} \lambda_0 z v^2 &=& 0.
\label{mincon}
\end{eqnarray}
Therefore, $v^2 \simeq -2m_0^2/\lambda_1$ as usual, but $u \simeq \lambda_0 
z v^2/2m_\xi^2$, with $u,z \ll v$.  The small VEV $u$ of the triplet, 
which gives masses to the neutrinos via \Eq{yuk}, is proportional to the 
value of $z$ and inversely proportional to the square of the Higgs triplet 
mass, i.e. $m_\xi^2.$  Thus the smallness of the singlet VEV $z$ together
with the possible suppression by other free parameters of the model
should ensure the correct order of magnitude for $u$ as determined by the
scale of the neutrino masses.

Solving \Eq{mincon} for the parameters $m_0^2$ and $m_\xi^2$, 
the mass-squared matrix of the neutral scalar fields in the 
$(H,\zeta)$ basis is given by
\begin{equation}
{\cal M}^2_{S} = 
\left[ \begin{array} {c@{\quad}c} \lambda_1 v^2 & \lambda_3 u v - 
\lambda_0 z v \\ \lambda_3 u v - \lambda_0 z v &  
{1 \over 2}  \lambda_0  v^2 z/u  + \lambda_2 u^2 \end{array} \right].
\label{ms}
\end{equation}
Since $u,z \ll v$, the fields $H$ and $\zeta$ are almost exact mass
eigenstates.  Thus $H$ behaves just like the SM Higgs boson and $\zeta$ 
is a heavy neutral scalar boson of mass $\simeq m_\xi$. 

For the pseudoscalar fields, the mass-squared matrix
in the  basis $(z \varphi, u \eta, v \theta)$ is given by
\begin{equation}
{\cal M}^2_{PS} = {1 \over 2} \lambda_0 z v^2 u \left[ \begin{array} 
{c@{\quad}c@{\quad}c} {1/z^2} & -{1 /z u} & {2 / z v} \\ 
-{1 /z u} & {1 / u^2} & -{2 / u v} \\ {2 / z v} & 
-{2/ u v} & {4 / v^2} \end{array} \right].
\label{mps}
\end{equation}
This mass matrix can be diagonalized by the orthogonal matrix
\bea
U=U_J\,U_G\,,
\label{U}
\eea
where
\bea
U_G=\frac{1}{\sqrt{v^2+ 4u^2}}
\left(
\begin{array}{ccc}
1 & 0 & 0 \\
0 & 2 u & v \\
0 & -v & 2u
\end{array}
\right)\,,
\eea
and
\bea
U_J=\frac{1}{\sqrt{u^2v^2/(v^2+ 4u^2)+ z^2}}
\left(
\begin{array}{ccc}
z & 0 & -uv/\sqrt{v^2+ 4u^2} \\
0 & 1 & 0 \\
uv/\sqrt{v^2+ 4u^2} & 0 & z
\end{array}
\right)\,.
\eea
Thus the physical mass eigenstates can be found from
\bea
\left(
\begin{array}{c}
J^0 \\ G^0 \\ \Omega^0
\end{array}
\right) = U 
\left(
\begin{array}{c}
z\varphi \\ u\eta \\ v\theta
\end{array}
\right)\,,
\label{mvsg}
\eea
where $G$ is the Goldstone mode giving mass to the $Z$ boson, 
$J^0$ is the physical massless Majoron, and $\Omega^0$ is the physical 
massive pseudoscalar boson which is mostly triplet and is the partner of 
$\zeta$. The factorization of $U$ in \Eq{U} is particularly useful
since it allows the immediate recognition of the $Z$-boson longitudinal 
component given explicitly by
\bea
G^0=\frac{v^2 \theta + 2 u^2 \eta}{\sqrt{v^2 + 4 u^2}}\,,
\eea
as well as the physical Majoron
\begin{equation}
J^0 = {(v^2 + 4 u^2) z^2 \varphi + v^2 u^2 \eta - 2 u^2 v^2 \theta \over 
\sqrt {z^2(v^2 + 4 u^2)^2 + u^2 v^4 + 4 v^2 u^4}}.
\label{j0}
\end{equation}
The massive combination of $(z \varphi, u \eta, v \theta)$ is of course 
\bea
\Omega^0=\frac{\varphi - \eta + 2 \theta}{\sqrt {z^{-2} + u^{-2} + 4 v^{-2}}}
\,,
\eea
with its mass-squared given by 
\bea
M^2_\Omega= \frac{1}{2}\lambda_0\left( 
v^2\frac{z}{u} + 4 uz + v^2\frac{u}{z}
\right)\,.
\eea
Notice that $\Omega^0$ is almost degenerate in mass with $\zeta$ as expected.

At this point a comment is in order. Notice that the massive Higgs singlet 
propagating in the bulk, i.e. $\rho,$ is completely decoupled from the 
Higgs fields living in our 3-brane. The only connection between these
two sectors is due to their couplings to the Majoron $J^0.$

Similarly we find the masses of the charged Higgs bosons.
The singly-charged Higgs mass-squared  matrix in the basis $(\phi^+, \xi^+)$
is found to be
\bea
{\cal M}^2_+=
\left(\lambda_0 z + \frac{1}{2}\lambda_5 u\right)\left[
\begin{array}{cc}
 u  &  v/\sqrt{2}\\
 v/\sqrt{2} & v^2/(2u)
\end{array}
\right]\,.
\eea
The longitudinal component of $W^+$ is easily found to be
\bea
G^+=\frac{ v\,\phi^+ -\sqrt{2}u\,\xi^+}{\sqrt{v^2+2u^2}}\,,
\eea
while the massive physical charged Higgs boson is orthogonal to that,
\bea
h^+=\frac{\sqrt{2}u \,\phi^+ + v\, \xi^+}{\sqrt{v^2+2u^2}}\,,
\eea
with mass squared
\bea
M^2_{h^{+}}=\frac{1}{2}\left(\lambda_0 \frac{z}{u} +
\frac{1}{2}\lambda_5 \right)\left(v^2 + 2 u^2\right)\,.
\eea
The would-be Goldstone boson is predominantly doublet while the physical 
charged Higgs boson is predominantly triplet.

Finally, the mass of the doubly-charged Higgs boson $\xi^{++}$ is given by
\bea
M^2_{\xi^{++}}=\frac{1}{2}\left(\lambda_0 \frac{z}{u} +\lambda_5 \right)v^2
+ 2\lambda_4 u^2\,.
\label{m++}
\eea
Therefore $M^2_{\xi^{++}}-M^2_{h^+} \approx M^2_{h^+}- M^2_{\zeta} \approx
\lambda_5 v^2/4$ as expected.

\section{Higgs Phenomenology at Colliders}

\subsection{Neutral sector}

In hadron and lepton colliders, the neutral Higgs bosons can be 
produced via the gauge interactions in the Drell-Yan and Bjorken
processes mediated only by the $Z$ boson. The production follows by the 
kinematically allowed decays 
\bea
\zeta^0,\;\Omega^0 & \to &  \nu\nu, \bar \nu \bar \nu\,, \label{01} \\
\zeta^0,\;H^0 & \to & Z J^0 \,, \label{02} \\
\zeta^0,\;H^0 & \to & J^0 J^0 \,, \label{03} \\
\zeta^0 & \to & H^0 H^0 \,, \label{04} \\
\Omega^0 & \to & H^0 J^0 \label{05} \,.
\eea
Here the first decays \rfn{01} to neutrinos come from the Yukawa 
interaction of \Eq{yuk} and their widths depend on the magnitudes of the 
Yukawa couplings. The others follow from the scalar self-interactions. 
The decays \rfn{04} and \rfn{05} are suppressed because $u,z \ll v$. 
There are of course also the well-known SM decays of $H^0$, but we do not 
discuss them here.  Thus the processes involving the new neutral scalar 
bosons $\zeta^0$ and $\Omega^0$ are practically invisible and are not of 
great phenomenological interest at collider experiments.  On the other hand, 
as we show below, they do constrain the allowed parameter space of our model.

The couplings of the Majoron to the other Higgs bosons follow from the 
kinetic-energy terms involving $\phi^0$ and $\xi^0.$ 
These are given by
\begin{equation}
\partial_\mu \bar \phi^0 \partial^\mu \phi^0 + \partial_\mu \bar \xi^0 
\partial^\mu \xi^0 = {1 \over 2} (\partial_\mu H)^2 + {1 \over 2} (H+v)^2 
(\partial_\mu \theta)^2 + {1 \over 2} (\partial_\mu \zeta)^2 + {1 \over 2} 
(\zeta + u)^2 (\partial_\mu \eta)^2.
\end{equation}
Reversing \Eq{mvsg} and substituting the fields into the interaction 
terms above, we find a term involving $(\partial_\mu J)^2$ 
given by
\begin{equation}
{(4 v u^4 H + u v^4 \zeta) (\partial_\mu J)^2 \over z^2 (v^2 + 4 u^2)^2 
+ u^2 v^4 + 4 v^2 u^4} \,.
\label{zjj}
\end{equation}
Because $u,z\ll v$, the coupling $H (\partial_\mu J)^2$ is suppressed
by the small factor $(u/v)^3$ so the decay $H \to JJ$ is completely 
negligible.  However, the coupling $\zeta (\partial_\mu J)^2$ is not 
suppressed.  To the contrary, the decay of $\zeta$ into two massless 
Majorons in \rfn{03} is enhanced by the large mass of $\zeta$ 
($\partial_\mu \to M_\zeta$ in the calculation). Indeed
\bea
\Gamma(\zeta\to JJ)\approx \frac{1}{64\pi} \frac{M^3_\zeta u^2}{(u^2+z^2)^2},
\eea
which implies that in order for the $\zeta$ width not to exceed its mass,
$z$ is required to be at least of order MeV, i.e. much larger than the 
scale of the neutrino masses.  Therefore we must have $u\ll z\ll v$. 

As for \rfn{02}, these processes come from the terms $2 g_Z \zeta Z_\mu 
\partial^\mu (u \eta)$ and $g_Z H Z_\mu \partial^\mu (v \theta)$.  They are 
suppressed by $u/z$ and $2u^2/zv$ respectively and are thus negligible.

\subsection{Singly-charged sector}

In addition to its coupling to the $Z$ boson, the charged physical Higgs boson 
$h^+$ (which is a triplet up to the negligible $u/v$ component of the doublet) 
couples also to the photon.  If kinematically accessible, it can be 
pair-produced via the Drell-Yan process at hadron and lepton colliders. Its
kinematically allowed decays are
\bea
h^+ & \to & l^+ \bar \nu \,, \label{+1} \\
h^+ & \to & W^+ J^0 \,, \label{+2} \\
h^+ & \to & W^+ \Omega^0 \,, \label{+3} \\
h^+ & \to & W^+ \zeta^0 \,. \label{+4} 
\eea
Here the last two decays may not be kinematically allowed (see the previous 
section).  Even if allowed, the decays \rfn{+3} and \rfn{+4} will be 
suppressed by phase space.  The decays \rfn{+1} and \rfn{+2} are always 
allowed kinematically.  However, because the triplet component in the 
Majoron is suppressed by the factor $u/z$ (see \Eq{j0}), the decay \rfn{+2} 
may not compete with \rfn{+1} unless the Yukawa coupling of the latter is 
very small.   Therefore, the best candidate for the $h^+$ decay channel is 
likely to be the decay \rfn{+1} induced by the Yukawa Lagrangian of \Eq{yuk}.  

The expected experimental signature of the process $pp\to h^+h^-$ at the 
LHC is two hard oppositely charged leptons plus large missing energy 
carried away by the neutrinos in the decay \rfn{+1}.  If the decay 
\rfn{+1} is suppressed by very small Yukawa couplings, the other decays 
\rfn{+2}, \rfn{+3}, \rfn{+4} may also be relevant.  However, the leptons 
coming from the $W^+$ decays are softer and may be discriminated from the 
decay products of \rfn{+1}.

\subsection{Doubly-charged sector}

\begin{figure}[t]
\centerline{
\epsfxsize = 0.45\textwidth \epsffile{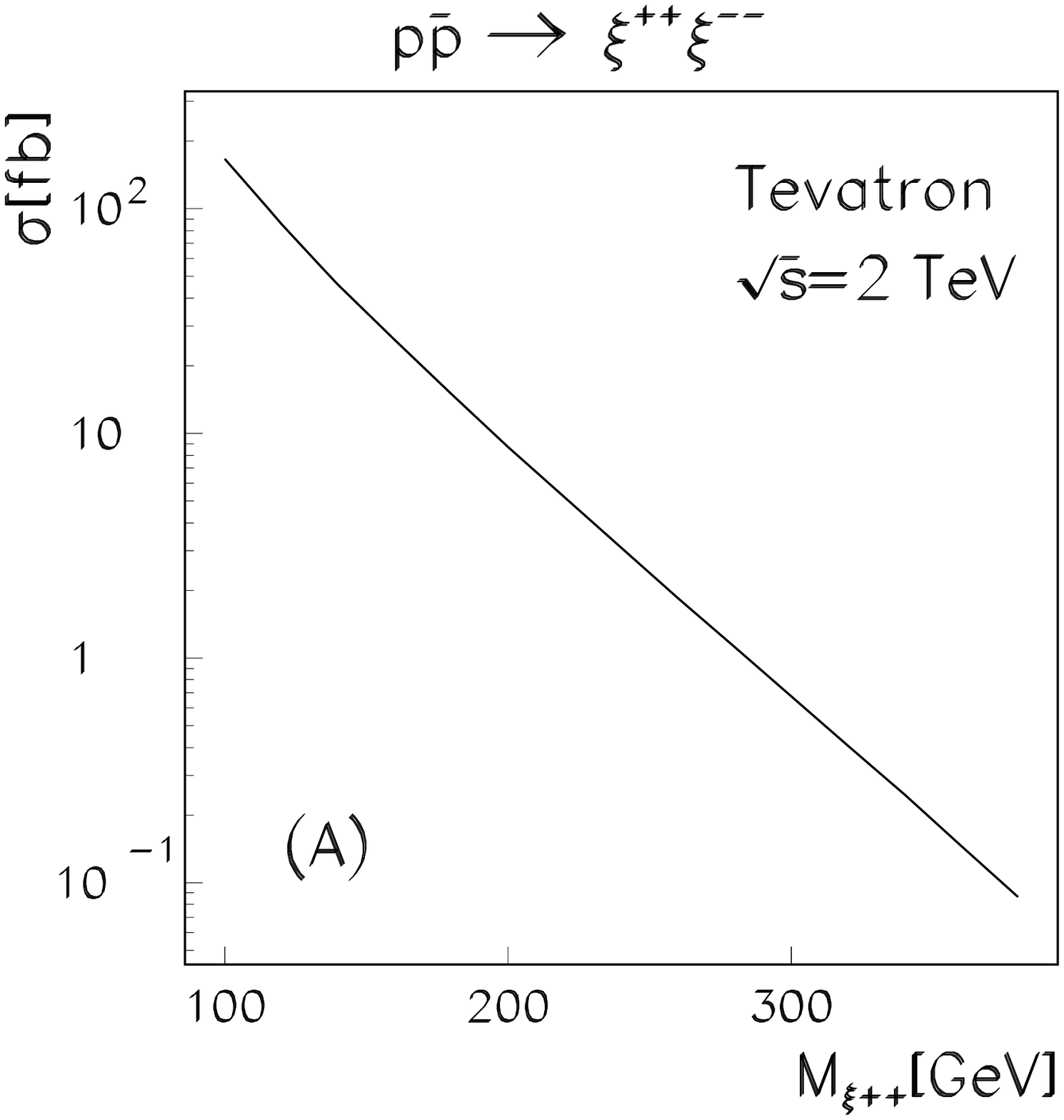}
\hfill                                                              
\epsfxsize = 0.45\textwidth \epsffile{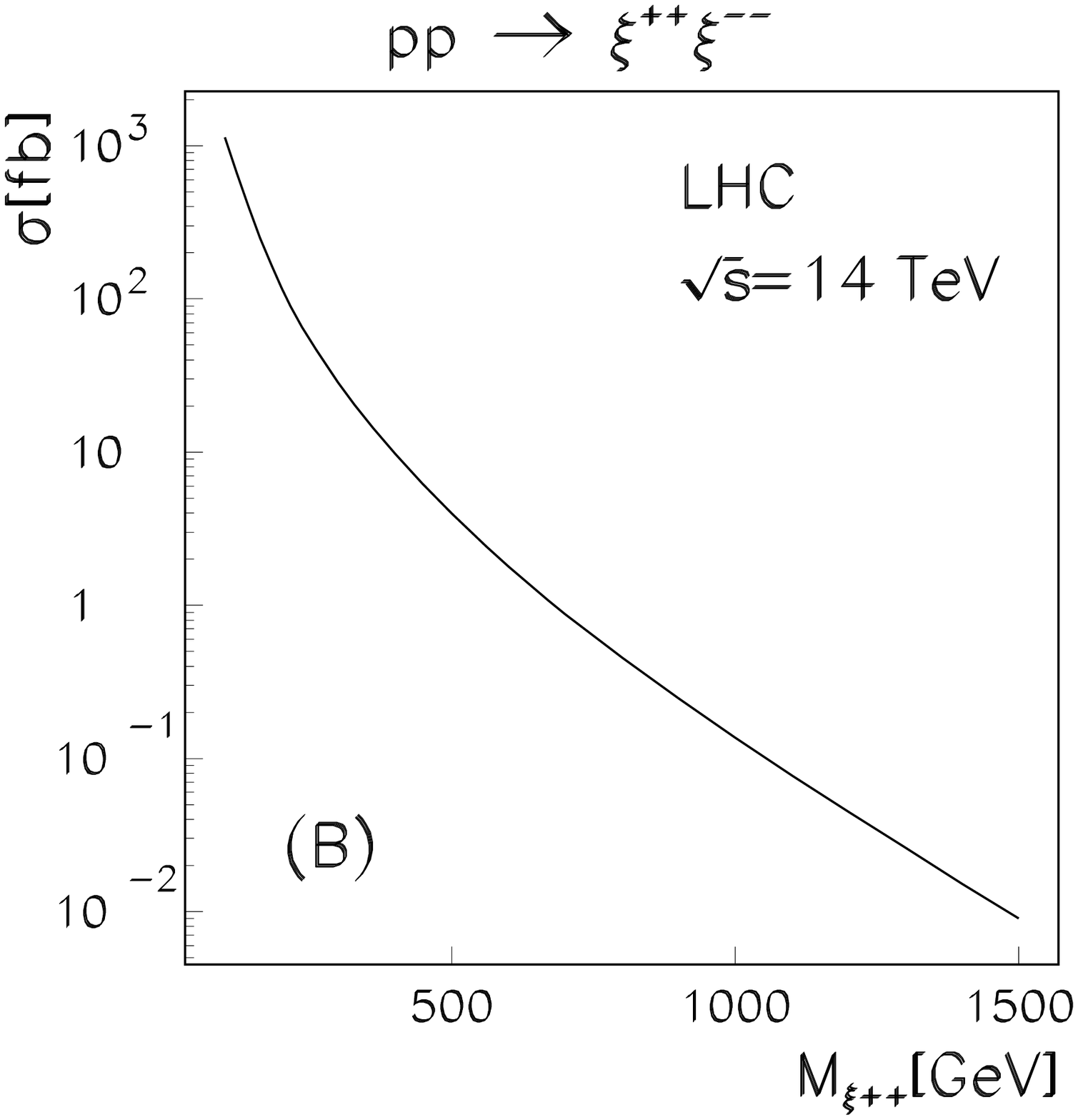} 
}
\caption{Cross section of $\xi^{++}\xi^{--}$ Drell-Yan pair production 
at Tevatron (A) and LHC (B). }
\label{fig:ppdd}
\end{figure}

The production of doubly charged Higgs bosons at future colliders 
offers background-free and complete experimental tests of our model of
neutrino masses.  The only pair-production mechanism of $\xi^{++}$ at 
the LHC and Tevatron is the Drell-Yan process mediated by s-channel photon 
and $Z$ exchange \cite{hmpr}.  Thus the production rate is enhanced by the 
double charge of $\xi^{++}$ and is uniquely determined by the gauge couplings. 
At the parton level, the differential cross section of the process
\bea
\bar f f \to \xi^{++} \xi^{--}\,,
\eea
where $f=u,d$ quarks, is given by 
\bea
\frac{d\sigma}{dt}=\frac{e^4}{48\pi s^2} {\cal M}^2\,,
\eea
where $s,t$ are the kinematical invariants and the squared amplitudes 
${\cal M}^2={\cal M}_1^2+{\cal M}_2^2+{\cal M}_{12}^2$ read
\bea
{\cal M}_1^2 &=& -\frac{8 Q_f^2}{s^2}
\left[(t-M^2_{\xi^{++}})^2 + st \right]\,, \\
{\cal M}_2^2 &=& -\frac{(1+X_f^2)}{8(s-M_Z^2)^2}
\left(\frac{1-2\sin^2\theta_W}{\sin^2\theta_W\cos^2\theta_W}\right)^2
\left[(t-M^2_{\xi^{++}})^2 + st \right]\,, \\
{\cal M}_{12}^2 &=& -\frac{2Q_fX_f}{s(s-M_Z^2)}
\left(\frac{1-2\sin^2\theta_W}{\sin^2\theta_W\cos^2\theta_W}\right)
\left[(t-M^2_{\xi^{++}})^2 + st \right]\,.
\eea 
Here  $Q_f=2/3,$ $X_f=1-(8/3)\sin^2\theta_W$ for $f=u,$ and 
$Q_f=-1/3,$ $X_f=-1+(4/3)\sin^2\theta_W$ for $f=d.$  
To obtain the cross 
section of $pp,p\bar p\to\xi^{++}\xi^{--}$ at the LHC and Tevatron, 
we have calculated the 
subprocesses involving the $u,\,\bar u$ and $d,\,\bar d$ collisions and 
convoluted them over the parton distributions given by the default set of the 
CERN library package PDFLIB \cite{pdflib}.  The total cross section 
as a function of the triplet mass $M_{\xi^{++}}$ is plotted in 
Fig.~\ref{fig:ppdd}.

Once produced, $\xi^{++}$ will decay via one of the following channels
\bea
\xi^{++} & \to & l^+ l^+ \,, \label{++1} \\
\xi^{++} & \to & W^+ W^+ \,, \label{++2} \\
\xi^{++} & \to & h^+ W^+ \,. \label{++3} 
\eea
Because  
$\langle \xi\rangle=u$ is tiny, the decay branching fraction of $\xi^{++}\to 
W^+W^+$ is negligible.  The decay \rfn{++3} may not be allowed kinematically 
and is suppressed by phase space in any case.  Thus the only unsuppressed 
decay channels in our scenario are $\xi^{++}\to l_i^+l_j^+$ with the partial 
rates 
\bea
\Gamma_{ij}=|f_{ij}|^2 \frac{M_{\xi^{++}}}{4\pi}\,, 
\eea
for $i\neq j,$ and 1/2 smaller for $i=j.$  This same-sign dilepton signal at 
the invariant mass of $\xi$ is very distinctive at the LHC or Tevatron 
because it is completely \underline {background-free}.  Assuming the total 
integrated luminosity of the LHC to be 1000 $fb^{-1}$ (10 $fb^{-1}$ at 
the Tevatron), the reconstruction efficiency of the event to be 10\%, 
and the predicted average of $N=-\ln(1-p)$ Poisson distributed 
events to provide a discovery, the cross sections in Fig.~\ref{fig:ppdd} 
imply at $p=95\%$ confidence level that $M_{\xi^{++}}\lsim 1.2$ TeV 
($M_{\xi^{++}}\lsim 300$ GeV) can be probed at the LHC ( Tevatron).  
Its decay branching fractions 
will then determine $|f_{ij}|$, i.e. the magnitude of each element of the 
neutrino mass matrix up to an overall scale factor.  This is the only model 
of neutrino masses which has the promise of being verified directly 
\cite{ma00} from collider experiments without involving other theoretical
assumptions which must be tested elsewhere.

Complementary measurements of $|f_{ij}|$ are also provided by the 
resonant processes $e^-e^-$ $(\mu^-\mu^-) \to l_i^-l_j^-$ at a future 
Linear Collider and/or Muon Collider. 
The $M_\xi^{++}$ reach in these colliders extends up to the collision 
energies, which may be as high as 4 TeV.  The sensitivity to $|f_{ij}|$ 
depends on the beam properties of the machines.  The detailed estimate 
in Ref.~\cite{cr} implies that $|(f\cdot f^*)_{ij}|\gsim 10^{-8}$ can be
probed in these processes.

\section{Neutrino Masses and Predictions for 
Lepton-Flavour Violating Processes }

In our model the Majorana neutrino mass matrix follows from \Eq{yuk}
and is given by
\bea
({\cal M}_\nu)_{ij}=2 f_{ij} \langle\xi\rangle.
\eea
The VEV of the triplet $ \langle\xi\rangle=u/\sqrt{2}$ should be 
derived from the minimization conditions of \Eq{mincon} which are
nonlinear in $u.$ Notice, however,  that $u$ identically vanishes if
$\lambda_0 z\to 0.$ Because of the hierarchy among the VEVs,
$u\ll z\ll v,$ we are allowed to make good approximations to relate
the value of $u$ to other VEVs and to physical Higgs-boson masses.
Let us choose the $\xi^{++}$ mass to be the physical parameter. Then
it follows from \Eq{m++} that
\bea
u\approx \frac{1}{2} \lambda_0 z \frac{v^2}{M^2_{\xi^{++}}}\,,
\eea
and the neutrino mass matrix takes the form
\bea
({\cal M}_\nu)_{ij}\approx \frac{1}{\sqrt{2}} f_{ij} \lambda_0 z 
\frac{v^2}{M^2_{\xi^{++}}}\,.
\label{mn}
\eea
For our phenomenological study of neutrino masses we treat 
the combination $\lambda_0 z$ as  a small free parameter;
the correct order of magnitude of the VEV $z$ is given 
by \Eq{chivev}.
As mentioned before, because the neutrino mass matrix \rfn{mn} is 
proportional to the Yukawa matrix $f_{ij}$, the branching fractions of 
the $\xi^{++}$ decays determine the neutrino masses up to the overall scale
which should be fixed from other experiments. On the other hand,
the existing neutrino data may already be used to make predictions for the
rare unobserved lepton flavour violating processes induced by $f_{ij}$
in our model.

Consider now a phenomenological hierarchical neutrino mass matrix consistent 
with the atmospheric and solar neutrino results \cite{neutrino,valle}:
\bea
{\cal M}_\nu  = m \pmatrix{ 0 & b & -b x \cr b & x^2+a & x-ax \cr -b x
& x-ax & 1+a x^2} ,
\label{mnmatr}
\eea
where $m$ is the normalization mass and $0.67 < x < 1$ determines the 
$\nu_\mu \to \nu_\tau$ mixing as required by the atmospheric neutrinos.  
The three solutions of the solar neutrino problem correspond to 
\begin{itemize}
\item[(i)] large-angle matter-enhanced oscillations: $a=0.02,~b=0.4$; 
\item[(ii)] small-angle matter-enhanced oscillations: $a=0.04;
~b=0.003$; and 
\item[(iii)]  vacuum oscillations: $a = 0.002,~b=0.012$.
\end{itemize}
While all these solutions to neutrino anomalies are still allowed,
the new global fits of neutrino data \cite{valle}, which include the recent
Super-Kamiokande data \cite{sknew}, clearly prefer the large angle
matter-enhanced solution (i).

Given the pattern of $f_{ij}$ via the neutrino mass matrix of 
Eq.~\rfn{mnmatr}, lepton-flavour violation through $\xi$ exchange may be 
observable at low energies. The processes most sensitive to the new 
flavour-violating physics are the decay $\mu\to e\gamma$ and $\mu-e$ 
conversion in nuclei.  Planned experiments will reach the sensitivity of 
$10^{-16}$ for $\mu-e$ conversion in aluminium \cite{meco} and $10^{-14}$ for 
$\mu\to e\gamma$ \cite{mueg}. 
Because of the off-shell photon exchange,
the amplitude of $\mu-e$ conversion in nuclei 
is enhanced by $\ln(M^2_{\xi^{++}}/m_\mu^2)$ compared to that of $\mu\to e
\gamma$ \cite{rs}.  Therefore we expect that the former process is more
sensitive to the existence of our neutrino-mass-giving triplet than the latter.

The matrix element of photonic conversion 
is given by 
\bea
{\cal M}=(4\pi\alpha/q^2) j^{\mu} J_{\mu}\,,
\eea
where $q$ is the 
momentum transfer with $q^2 \approx -m^2_\mu,$ $J$ is the hadronic current, 
and
\bea
j^{\lambda} &=& {\bar u}(p_e) \left[ \; 
\left (f_{E0} + \gamma_5 f_{M0}\right) \gamma_{\nu}  \left (
g^{\lambda \nu} - \frac {q^{\lambda} q^{\nu}}{q^2}\right ) +
(f_{M1} + \gamma_5 f_{E1})\; i\; \sigma^{\lambda \nu} \frac{q_{\nu}}{m_{\mu}} 
\right] u(p_{\mu})\,
\label{j1}
\eea
is the usual leptonic current.  
The coherent $\mu-e$ conversion ratio in nuclei 
is given by
\beq
R_{\mu e}=\frac{8\alpha^5\,m_\mu^5\,Z^4_{eff}\,Z\,|\overline{F_p}(p_e)|^2}
{\Gamma_{capt}}
 \frac{\xi_0^2}{q^4}\, ,
\label{mecrate}
\eeq
where $\xi_0^2=|f_{E0}+f_{M1}|^2+|f_{E1}+f_{M0}|^2$, and for $^{13}$Al, 
$Z^{Al}_{eff}=11.62,$  $\overline{F_p}^{Al}(q)=0.66$, and
$\Gamma_{capt}^{Al}=7.1\times 10^5$~s$^{-1}$ \cite{chiang}.
We calculate the form factors induced by the one-loop diagrams involving
$\xi^{++}$ and obtain  
\bea
&& f_{E0} = f_{M0}= \sum_l {f_{\mu l} f^*_{l e} \over 24\pi^2} [4s_l+rF(s_l)], 
\label{f0}
\\
&& f_{M1} = -f_{E1}= \sum_l {f_{\mu l} f^*_{l e} \over 24\pi^2} s_\mu,
\label{f1}
\eea
where $$F(s_l) =
\ln s_l+\left(1-\frac{2s_l}{r}\right)\sqrt{1+\frac{4s_l}{r}}
\ln\left[\frac{\sqrt{r+4s_l}+\sqrt{r}}{\sqrt{r+4s_l}-\sqrt{r}}\right],
$$
and $r=-q^2/M_{\xi^{++}}^2$, $s_l=m_l^2/M_{\xi^{++}}^2,$ $l=e,\,\mu,\,\tau$. 
In the interesting limit $s_l\to 0$, we get $F(s_l)\to \ln r$ which implies
the logarithmic enhancement of the form-factors $f_{E0}$ and  $f_{M0}.$ 
Notice that all the form factors in \Eq{j1} contribute to the 
$\mu-e$ conversion rate \rfn{mecrate}. However, it is well known that 
the decay $\mu\to e\gamma$
is induced only by the form-factors $f_{E1}$ and  $f_{M1}.$ 
Its branching ratio is given by 
\bea
R_{\mu\to e\gamma}=\frac{96 \pi^3\alpha}{G_F^2m_\mu^4}
\left( |f_{M1}|^2 + |f_{E1}|^2 \right) \,,
\eea
where $\alpha=1/137$ and $G_F$ is the Fermi constant.
\begin{figure}[t]
\centerline{
\epsfxsize = 0.5\textwidth \epsffile{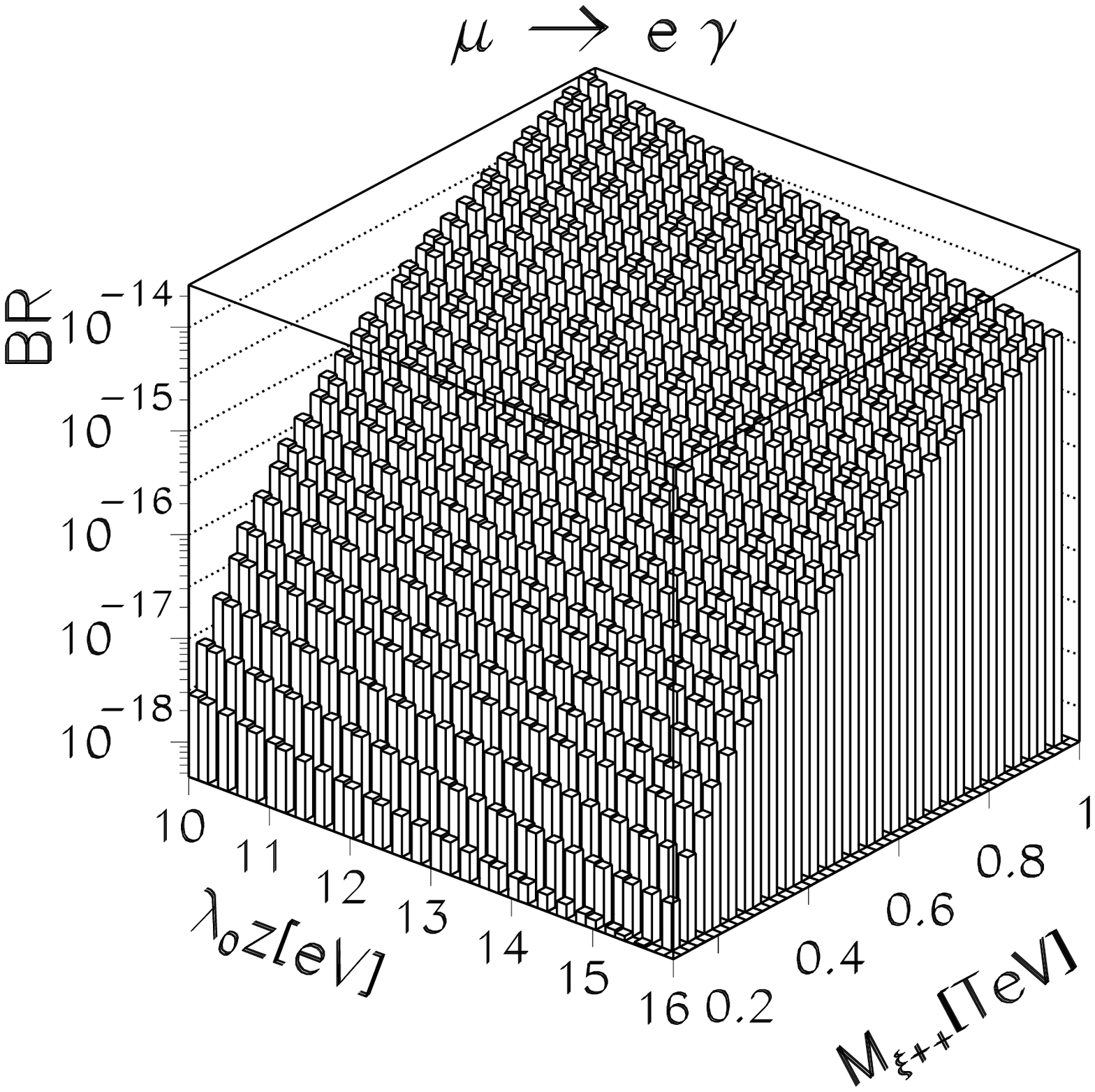}
\hfill 
\epsfxsize = 0.5\textwidth \epsffile{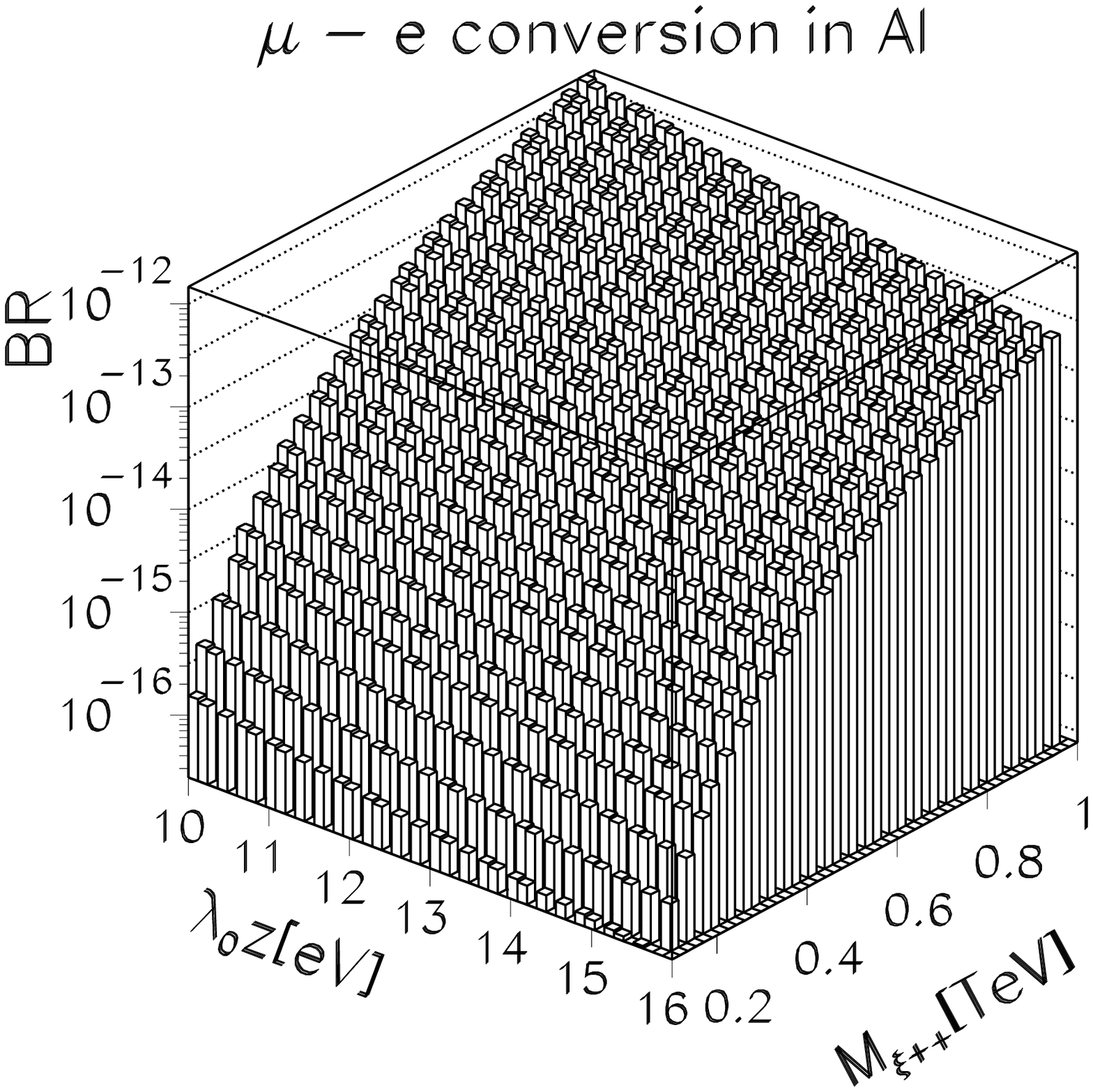}
}
\caption{Rates of $\mu\to e\gamma$ and  
$\mu-e$ conversion in $^{13}$Al against the $\xi^{++}$ mass
and the free parameter $\lambda_0 z$, assuming large-angle 
matter-enhanced solution to the solar neutrino problem.  }
\label{fig:muec}
\end{figure}

For numerical estimates we assume $m=0.03$ eV and $x=0.9$ in \Eq{mnmatr} 
and the currently most-favoured large-angle matter-enhanced 
oscillation solution (i) to the solar neutrino problem.
In Fig.~\ref{fig:muec} we plot the branching ratio of the decay 
$\mu\to e\gamma$ and the ratio of $\mu-e$ conversion in 
aluminium as a function of the mass $M_{\xi^{++}}$ and the free parameter
$\lambda_0 z$. The behaviour of these 
ratios can be understood from \Eq{mn}: a fixed neutrino mass implies 
$f \propto M_{\xi^{++}}^2 $ and $f \propto 1/(\lambda_0 z).$ 
Notice the complementarity 
of collider and $\mu-e$ conversion experiments.  For small $M_{\xi^{++}}$, 
$R_{\mu e}$ is suppressed while the collider cross section is kinematically 
enhanced, and vice versa.   
\begin{figure}[t]
\centerline{
\epsfxsize = 0.6\textwidth \epsffile{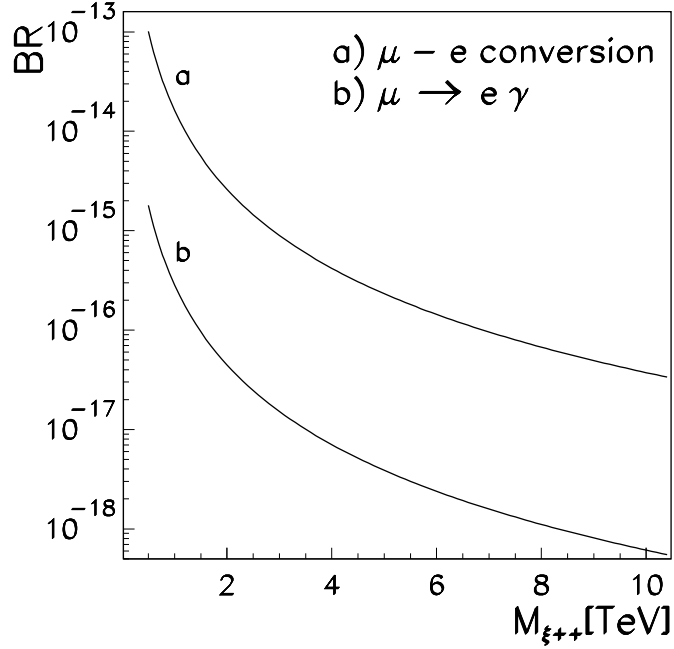}
}
\caption{Rates of $\mu\to e\gamma$ and  
$\mu-e$ conversion in $^{13}$Al against the $\xi^{++}$ mass.
We assume that at $M_{\xi^{++}}=0.5$ TeV  
$\lambda_0 z=10$ eV, and it scales according
to \Eq{chivev} for $n=3$ extra dimensions.  }
\label{fig:mue2}
\end{figure}

For the same parameters $\lambda_0 z$ and $M_{\xi^{++}}$, the numerical 
values of $R_{\mu e}$ are almost two orders of magnitude larger than the 
ones of $R_{\mu\to e\gamma}.$  There are two reasons. First, $R_{\mu e}$ 
is enhanced by the large logarithm in \Eq{f0}. Second, there is a deep 
cancellation between the triplet Yukawa couplings in
\Eq{f1}  which follows from the structure of the mass matrix \Eq{mnmatr}
[the minus sign in the (13), (31) entries].
In \Eq{f0}, however,  the cancellation does not occur because the 
flavour-dependent multiplicative function gives different 
weights to different contributions in the sum. Hence $R_{\mu\to e\gamma}$
(which depends only on $f_{M1},$ $f_{E1}$) is further diminished compared to
$R_{\mu e}.$ One should remember
that the sensitivity of the planned $\mu-e$ conversion experiments
is at least two orders of magnitude higher than the sensitivity of the
planned  $\mu\to e\gamma$ experiments. Therefore the  $\mu-e$ conversion 
experiments will probe the existence of the neutrino-mass-giving triplet 
to high mass scales while the $\mu\to e\gamma$ experiments will have just
a marginal chance to test this scenario. 

This can also be seen in Fig.~\ref{fig:mue2} where we plot the branching ratios
of $\mu\to e\gamma$ and $\mu-e$ conversion in Al against $M_{\xi^{++}}.$
In doing so, we assume that the triplet mass is equal to the 
fundamental scale of the theory, $M_{\xi^{++}}=M_*,$ and the VEV of the 
singlet $\chi$ (and thus $\lambda_0 z$) evolves according to \Eq{chivev}
for $n=3$ extra dimensions. We assume the initial value $\lambda_0 z=10$ eV
for $M_{\xi^{++}}=M_*=0.5$ TeV. For this choice of parameters, 
Fig.~\ref{fig:mue2} indicates that MECO will test our model up to
the scale of 7 TeV while the $\mu\to e\gamma$ rate is always
below the sensitivity of the currently planned experiments.

\section{Conclusions}

The neutrino-mass-giving Higgs triplet $(\xi^{++}, \xi^+, \xi^0)$ is 
proposed to be observable at future colliders with $m_\xi$ of order 1 TeV in 
a model where spontaneous lepton-number violation comes from a scalar 
bulk singlet $\chi$ in a theory of large extra dimensions.  We show how $\xi$ 
couples to the massless Majoron $J^0$ of this model and study the various 
decay modes of the Higgs triplet components.  The backgroundless decays 
$\xi^{++} \to l_i^+ l_j^+$ will determine directly the relative magnitudes 
of the \{$ij$\} elements of the neutrino mass matrix.  The decay $\xi^0 \to 
J^0 J^0$ puts a severe constraint on the parameter space of this model, making 
$\langle \chi \rangle$ of order 1 MeV in our brane.  Using present 
neutrino-oscillation data, we predict observable rates of $\mu-e$ conversion 
in nuclei while the planned  $\mu \to e \gamma$ experiments will have
a much smaller chance for a positive signal.

{\it Acknowledgement.} This work was supported in part by the U.~S.~Department 
of Energy under Grant No.~DE-FG03-94ER40837 and ESF under Grant No. 3832. 
We thank M. Hirsch for pointing out an error in the neutrino mass matrix 
in the first version of this paper.

\newpage
\bibliographystyle{unsrt}

\end{document}